\documentclass[twocolumn,pra,superscriptaddress]{revtex4}%
\usepackage{graphicx}
\usepackage{amsmath}
\usepackage{amssymb}
\usepackage[ansinew]{inputenc}
\usepackage{bm}\let\mathbf\bm %usage \bm{a}; \let\mathbf\bm replaces \mathbf by \bm. It produces bold italic vectors.
\newcommand{\figwidth}{0.9\columnwidth}

\begin{document}
\title{Energy absorption of a Bose gas in a periodically modulated optical lattice}
\author{A. Iucci}
\affiliation{Universit\'{e} de Gen\`{e}ve, DPMC, 24 Quai Ernest-Ansermet CH-1211 Gen\`{e}ve 4, Suisse}
\author{M.~A. Cazalilla}
\affiliation{Donostia International Physics Center (DIPC) Manuel de Lardizabal 4,
20018-Donostia, Spain}
\author{A.~F. Ho}
\affiliation{School of Physics and Astronomy, Birmingham
University, Edgbaston, Birmingham B15 2TT, U. K.}
\author{T. Giamarchi}
\affiliation{Universit\'{e} de Gen\`{e}ve, DPMC, 24 Quai
Ernest-Ansermet CH-1211 Gen\`{e}ve 4, Suisse}
%\date{\today}

\begin{abstract}
We compute the energy absorbed by a one dimensional system of cold
bosonic atoms in an optical lattice subjected to lattice amplitude
modulation periodic with time. We perform the calculation for the
superfluid and the Mott insulator created by a weak lattice, and
the Mott insulator in a strong lattice potential. For the latter
case we show results for 3D systems as well. Our calculations,
based on bosonization techniques and strong coupling methods, go
beyond standard Bogoliubov theory. We show that the energy
absorption rate exhibits distinctive features of low dimensional
systems and Luttinger liquid physics. We compare our results with
experiments and find good agreement.
\end{abstract}

\maketitle

Cold atoms provide a remarkable laboratory to study the physics of
strongly correlated quantum systems. Cold atomic gases loaded in
optical lattices~\cite{greiner_mott_bec,stoferle_tonks_optical}
and Feshbach resonances~\cite{inouye_feshbach_resonances_bosons}
allow for an unprecedented control of many parameters of the system, including
the interactions, both for
bosons and fermions. However,  the ability to
probe the properties of such systems, and in particular, to
measure momentum and frequency dependent correlation functions
still remains very limited. One of the most common experimental probe is
time-of-flight (TOF) imaging, which under most conditions give
access to the the momentum
distribution~\cite{richard_1dbec_momentum}. More recently other
spectroscopies such as
Bragg~\cite{stenger_bragg_spectroscopy,hagley_measurement_coherence_bec},
energy absorption rate
(EAR)~\cite{stoferle_tonks_optical,schori_absorption_optical},
radio frequency~\cite{regal_ultracold_molecules_fermi}, and
shot-noise~\cite{shot-noise} have been demonstrated. Thus in order
to properly characterize these systems it is of primary importance
to have a good understanding as well as theoretical predictions
for such experimental probes.

In a series of recent experiments, the Zurich
group~\cite{schori_absorption_optical,stoferle_tonks_optical}
measured the EAR of Bose gases of ultracold $^{87}$Rb atoms loaded
in optical lattices. The system was heated by periodically
modulating the lattice along one direction, and the energy
absorption rate as a function of the modulation frequency was
subsequently estimated from the width of the atom distribution
around zero momentum measured in TOF. Interestingly, of all the
system dimensionalities reported in~\cite{stoferle_tonks_optical},
the one-dimensional (1D) ones exhibit the broadest continuous
spectrum away from the Mott regime, which has no simple
explanation in standard Bogoliubov
theory~\cite{stoferle_tonks_optical,popov_functional_int}, and has
led some to consider other
schemes~\cite{rey_absorption_optical_numerics,vanoosten_bragg_mott,batrouni_dynamic_response,kramer_absorption_optical_bogoliubov}.

In this work, we show that treating the 1D interacting Bose gas in
the framework of the Luttinger liquid
(LL)~\cite{haldane_bosons,gogolin_1dbook,cazalilla_correlations_1d,giamarchi_book_1d}
correctly takes into account short distance correlations that are
neglected in the Bogoliubov theory, and leads naturally to a
continuum of excitations, hence a continuous absorption spectrum.
Furthermore, these correlations are responsible for the transition
to the Mott insulating (MI) phase for sufficiently strong
interactions.

In contrast to the broad spectra observed for the superfluid (SF),
in Refs.~\onlinecite{stoferle_tonks_optical}
and~\onlinecite{schori_absorption_optical} two much narrower peaks
were observed for the Mott phase. Here we also study this regime
and consider two physically distinct situations: First, a
sufficiently strongly interacting Bose gas in 1D, as in
Ref.~\onlinecite{kinoshita_tonks_continuous}, becomes Mott
insulating for a weak \emph{commensurate} periodic
potential~\cite{haldane_bosons,giamarchi_mott_shortrev,buchler_cic_bec}.
In this case, we present results for the EAR  spectrum obtained
using the form factors for the effective low-energy sine-Gordon
model~\cite{lukyanov_sinegordon_formfactors}. These results also
apply to the MI phase of the 1D Bose-Hubbard model near the SF-MI
transition. We find that the spectrum exhibits a finite threshold,
corresponding precisely to the Mott gap. Second, we consider a
very deep optical lattice, so that hopping is strongly suppressed
and the Mott gap is large. This situation is well described by the
Bose-Hubbard model, for which we have computed in $1D$, the shape
of the lowest excitation peak in the absorption spectrum, which
occurs at an excitation frequency $\omega \approx U/\hbar$, $U$
being the on-site boson-boson repulsion. We also give the results
for the peak width in higher dimensions.

 Consider a system of interacting bosons in a lattice created
by an optical potential $V(x,y,z) = V_{0x} \sin^2 (k x) + V_{0 y}
\sin^2(ky) + V_{0z} \sin^{2}(kz)$, where the wave vector is
$k=2\pi/\lambda$, $\lambda$ being the laser wavelength. We first
consider an optical lattice that is very deep in the $y$ and $z$
direction (\emph{e.g.} $V_{0x} \ll V_\perp \equiv V_{0y} =
V_{0z}$). Atoms accumulate in the minima of this potential, where
they experience strong transverse confinement, thus forming 1D gas
tubes with a weaker periodic potential along the axis. For large
enough $V_{0y}$ and $V_{0z}$ each 1D system becomes isolated from
each other~\cite{ho_deconfinement_coldatoms}: this is the 1D
limit.

To obtain analytical results, we neglect trapping and finite size
effects~\footnote{\label{fn:confining}The trapping will lead to a
coexistence of the Mott phase with a small superfluid fraction at
the edges and near the center of the
trap~\cite{kollath_dmrg_bose_hubbard_trap}. This will contribute
to the energy absorption at low frequencies, but this part of the
spectrum is not well resolved  in the
experiments~\cite{schori_absorption_optical}. It also has marked
consequences for higher resonances in the spectrum (peak at
$\omega\approx 2U$) which is beyond the scope of the present
paper.}. If $V_{0x} \ll \mu$, where $\mu$ is the chemical
potential of the 1D interacting gas, the system in the presence of
the lattice is well described by the following sine-Gordon (sG)
model~\cite{haldane_bosons,giamarchi_mott_shortrev,buchler_cic_bec,giamarchi_book_1d}:
\begin{multline}\label{eq:solvableHamiltonian}
 H_{\rm eff} = \frac{\hbar v_{s}}{2\pi}\int dx\ \left[ K \left(\pi \Pi\right)^{2}
 + K^{-1}\left(  \partial_x\phi\right)^{2}  \right] \\
 + g_0 \int dx \cos (2\phi(x)),
\end{multline}
where $\phi(x)$ and $\Pi(x)$ are canonically conjugate fields;
$\phi$ represents density fluctuations, and
$\theta=\pi\int_{-\infty}^x\,dx'\Pi(x')$ corresponds to phase
fluctuations. $v_s$ is the speed of sound and the coupling to the axial potential
is  $g_0 \sim \rho_0
V_{0x} (V_{0x}/\mu)^{n_0-1}$, with $n_0$ the number of bosons per
potential well. $K$ is a
dimensionless parameter determined by the strength of the
boson-boson interactions and the linear density
$\rho_0$~\cite{cazalilla_correlations_1d,cazalilla_tonks_gases}:
the SF-MI transition occurs at the universal value $K_c = 2$ (in
terms of $\gamma = m g/\hbar^2
 \rho_0$, the dimensionless interaction strength of the
Lieb-Liniger model with $m$ the atom mass and $g$ proportional to
the scattering length, it corresponds to $\gamma\approx 3.5$).

A weak time dependent modulation of the
lattice~\cite{schori_absorption_optical,stoferle_tonks_optical}
$V_{0x} \to V_{0x} + \delta V_x \cos(\omega t)$, leads to a
perturbation that can be written in the above low-energy
description as $H'(t) = \delta V_x \: \cos(\omega t) \, {\cal O}$,
with ${\cal O} = f_0 \int dx \: \cos (2 \phi)$, where $f_0 =
dg_0/dV_{0x}$. Within linear response theory, the EAR per particle
at frequency $\omega$ is given by
\begin{equation} \label{eq:enabs}
 \dot{\epsilon}(\omega)=\frac{2 \delta V^2_{x}}{N}  \:  \: \omega \:
{\rm Im} \left[ - \chi_{\cal O}(\omega)  \right],
\end{equation}
where $\chi_\mathcal{O}(\omega)$ is the Fourier transform of the
retarded correlation function $ - i \hbar^{-1} \Theta(t) \langle
\left[\mathcal{O}(t), \mathcal{O}(0) \right] \rangle$, with
$\Theta$ the step function. Note that, in the bosonization
technique~\cite{cazalilla_correlations_1d,giamarchi_book_1d}, the
density
 $\rho(x) = \rho_0 - \partial_x \phi(x)/\pi + \cos \left[2
\phi(x) - 2\pi \rho_0 x\right] + \cdots$, and hence the EAR probes
the $q\approx 2 \pi \rho_0 = 2 n_0 k $ part of the excitation
spectrum. Standard Bogoliubov theory has no low energy excitation
near this momentum, hence no absorption, and one has to resort to
non-linear response to account for the observed
spectrum~\cite{kramer_absorption_optical_bogoliubov}. But in 1D,
the correct excitation spectrum has in fact a continuum at
$q\approx 2 \pi \rho_0$~\cite{lieb_excit,castroneto_exact_1d},
which is taken fully into account by the bosonization method.
%Thus we do find a finite and continuous absorption at low frequencies
%in linear response.

In the 1D SF (namely, LL) phase, where $K > 2$, the cosine term of
(\ref{eq:solvableHamiltonian}) is irrelevant and the system is
gapless. $\chi_{\mathcal{O}}(\omega)$ for $\hbar \omega \ll \mu$
can be obtained~\cite{iucci_absorption_long} by means of
bosonization
techniques~\cite{cazalilla_correlations_1d,giamarchi_book_1d}, and
we get:
\begin{equation}\label{result}
 \dot{\epsilon}(\omega)=\frac{A}{\hbar}\left( \frac{f_0 \delta V_{x}}{\rho_0} \right)^2 \:
 \left(\frac{\hbar\omega}{\mu}\right)^{2K-1},
\end{equation}
where $A$ is a  non-universal prefactor depending on the
microscopic details of the model (Bose-Hubbard or Lieb-Liniger).
Eq. (\ref{result}) directly shows that the continuum of low energy
excitations at $q\approx 2 \pi \rho_0$
 leads to this continuous absorption curve \footnote{If
the 1D potential $ V_{0x} \sin^2(k x)$ is not commensurate with
the density, then $\dot{\epsilon}(\omega>v_s \left|q\right|) =
{\cal A} \: \omega \left[ \omega^2 - v^2_s q^2 \right]^{K-1}$,
where $q = 2 n_0 k - 2\pi \rho_0$ is the incommensurability.}. As
$\hbar\omega \to \mu$, bosonization ceases to be valid. However,
standard sum rules for the density response
function~\cite{pitaevskii_becbook} imply that the integrated
absorption spectrum is finite, and thus the spectral weight must
decrease as $\omega \sim \mu/\hbar$ (or $\omega \sim J/\hbar$ for
the Bose-Hubbard model). One thus expects a rather broad spectrum
in 1D, as observed in the
experiments~\cite{schori_absorption_optical}.

\begin{figure}
 \begin{center}
 \includegraphics[width=\figwidth]{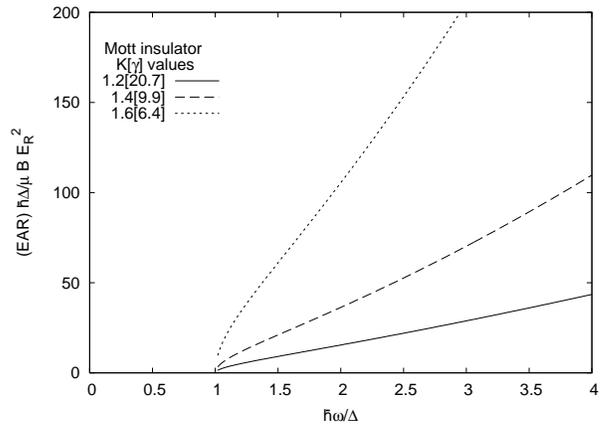}
 \caption{Energy absorption rate (EAR) for 1D interacting bosons
 in a weak optical lattice, for the MI phase (with small Mott gap)
 with different values of the parameter $K$ (in brackets are
 the corresponding values of $\gamma$~\cite{cazalilla_correlations_1d}).}\label{fig:weakLattice1D}
 \end{center}
\end{figure}

On the other hand, in dimensions higher than one, continuous
absorption must start from a finite frequency, as required by
Landau's criterion for superfluidity. We speculate that the
relatively broad absorption spectrum observed for the SF in the
``1D to 3D crossover'' regime~\cite{stoferle_tonks_optical} may be
due to the possible existence of a broad roton-like feature.
Since this particular lattice is highly anisotropic, a naive
extrapolation of the dimensional crossover theory of
Ref.~\cite{ho_deconfinement_coldatoms} suggests that at high
frequencies, the 1D-like excitation continuum persists, while at
low frequencies, the excitation is more 3D-like, with a broad
roton-like minimum being the vestige of the 1D continuum at
$q\approx 2\pi \rho_0$, but this minimum must not be at zero frequency to
satisfy Landau's criterion (see also
Ref.~\cite{nozieres_roton_unpublished}).

Next we turn to the MI phase in 1D, which occurs for sufficiently
repulsive interactions, $K < 2$. Now, the cosine term in
(\ref{eq:solvableHamiltonian}) leads to an excitation gap, with
gapped solitons and anti-solitons in the $1 < K < 2$ range of
interest here~\cite{gogolin_1dbook,giamarchi_book_1d}. Using the
form factor approach, and keeping only the one soliton- one
antisoliton contribution to the absorption, we
find~\cite{iucci_absorption_long}:
\begin{equation}
 \dot{\varepsilon}(\omega)=\frac{B}{\hbar}\left( \frac{f_0 \delta V_{x}}{\rho_0} \right)^2
\frac{\mu\,\Theta\left[(\hbar \omega)^{2}- \Delta^2 \right]}
 {\sqrt{(\hbar\omega)^{2}-\Delta^2}}
  \left\vert f\left[
 \theta_0(\omega)\right]\right\vert ^{2}, \label{eq:energyMottWeak1D}
\end{equation}
where $B$ is a non-universal prefactor, $\Delta = 2 M_s v^2_s$ is
the Mott gap,  and $M_s  \sim \mu (g_0/\rho_0\mu)^{1/(2-K)}/v^2_s$
($K <  2$) is the soliton gap (mass).
$\theta_0(\omega)=2\operatorname{arccosh}(\hbar\omega/\Delta)$ is
the relative rapidity of the soliton and the antisoliton, and
$f\left(\theta_0\right)
 =\frac{\sinh\theta_0}{\sinh\left(\frac{\theta_0+i\pi}{2\xi}\right)}\;
 e^{T\left(\theta_0\right)}$, where $T\left(\theta_0\right)=\int_{0}^{\infty}\frac{dt}{t}\frac{ \sinh
 ^{2}t\left(  1-\frac{i\theta_0}{\pi}\right) \sinh\left[
 t(\xi-1)\right]  }{\sinh(2t)\cosh (t)\sinh(t\xi)}$, is the (unnormalized) form factor of the
operator $\cos2\phi$~\cite{lukyanov_sinegordon_formfactors}, with
$\xi=K/(2-K)$. The EAR is plotted in
Fig.~\ref{fig:weakLattice1D} for various $K$, showing
the Mott gap to energy absorption. For  $\hbar \omega \gtrsim \Delta$
the EAR increases monotonically  in  a way slower than in the SF.

We now consider the case where the lattice potential is very deep
($\mu \ll V_{0x}$), for a $d$-dimensional hypercubic lattice. The
system  is then described~\cite{jaksch_bose_hubbard} by the
Bose-Hubbard model $H_{\rm BH} = H_J + H_U$, where
\begin{equation}\label{BH}
 H_{J} =    -\sum_{\mathbf{R},\mathbf{x}_{\alpha} } \frac{J_{\alpha}}{2}
 \: b^{\dag}_{\mathbf{R}} b^{\phantom{\dag}}_{\mathbf{R}+ \mathbf{x}_{\alpha}}, \,\,
 H_{U} =  \frac{U}{2} \sum_{\mathbf{R}} \left(\delta n^{\phantom{\dag}}_{\mathbf{R}} \right)^2
\end{equation}
%
%are the kinetic and interaction energy operators, respectively,
$b^{\dag}_{\mathbf{R}}$
creates  a boson at lattice site $\mathbf{R}$,
 $\mathbf{x}_{\alpha}$ are lattice vectors joining the
site $\mathbf{R}$ to  its two nearest neighbors along direction
$\alpha = 1, \ldots, d$; $\delta n^{\phantom{\dag}}_{\mathbf{R}} =
b^{\dag}_{\mathbf{R}} b^{\phantom{\dag}}_{\mathbf{R}} - n_0$. For
$V_{0\alpha}\gg E_R$ lattices, $J_{\alpha} =(8/\sqrt{\pi})E_R
(V_{0\alpha}/E_R)^{3/4} \exp[-2 (V_{0\alpha}/E_R)^{1/2}]$ and $U =
4\sqrt{2\pi}(a_s/\lambda)E_R \left(V_{0x} V_{0y} V_{0z}
/E^3_R\right)^{1/4}$ \cite{zwerger__JU_expressions}, with $a_s$
the scattering length, can be controlled by varying the laser
intensity $V_{0\alpha}$ (measured in units of the recoil energy
$E_R = \hbar^2 k^2/2m$). In one dimension ({\it i.e.} $d =1$, or
effectively when $J_1 \gg J_{\alpha}$, for $\alpha = 2, \ldots,
d$), the SF-MI transition occurs at $(U/J)_c =
1.92$~\cite{kuhner_bose_hubbard_critical_point}, while in a $d=3$
square lattice, $(U/J)_c = 5.8$, with $J=\sum_\alpha J_\alpha$.
Near the transition, on the MI side, the above description in
terms of the sG model~(\ref{eq:solvableHamiltonian}), still
applies~\cite{giamarchi_book_1d}. Thus, the EAR is also given
by~(\ref{eq:energyMottWeak1D}), and the Mott gap $\Delta$ is
exponentially small. As $U/J_1$ grows, eventually $\Delta \approx
U$.

Since $J_{\alpha} = J_{\alpha}(V_{0\alpha})$ and $U =
U(\{V_{0\alpha}\})$ are functions of the  optical potential
strength $V_{0\alpha}$, the modulation along one direction $V_{0x}
= V_{01} \to V_{0x} + \delta V_x \: \cos(\omega t)$ induces the
following perturbation to (\ref{BH})
\cite{reischl_temperature_mott}:
\begin{gather}\label{perturbation}
 H'(t)=\,\delta V_x F_U  \,H_{BH}+ \tilde{\mathcal{O}} \cos \omega t ,
\end{gather}
where $\tilde{{\cal O}} = - \frac{1}{2}\sum_{\mathbf{R}, \mathbf{x}_{\alpha}}
 \delta J_{\alpha} \, b^{\dag}_{\mathbf{R}}
b^{\phantom{\dag}}_{\mathbf{R} +  \mathbf{x}_{\alpha}}$, with
$\delta J_1= J_1 (F_J-F_U)\delta V_x$, $\delta J_\alpha = -
J_\alpha F_U\delta V_x$  ($\alpha > 1$), $F_{J} = d\ln
J_1/dV_{0x}$ and $F_{U} = d\ln U/dV_{0x}$. The first term in
(\ref{perturbation}) is $\propto H_{BH}$ and does not contribute
to the absorption.

\begin{figure}[h]
 \begin{center}
 \includegraphics[width=\figwidth]{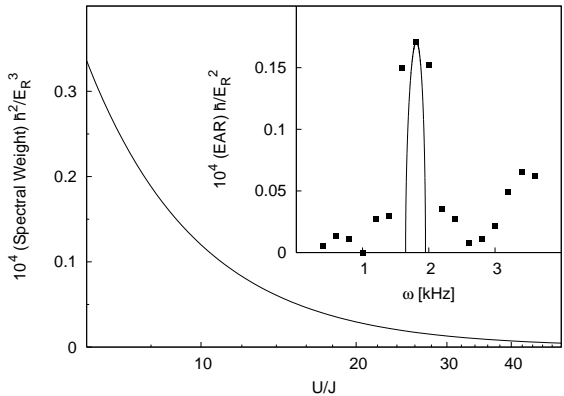}
 \caption{Spectral weight of the first resonance
  peak in the MI phase in 1D. In the inset we show the predicted 1D shape of the peak for $U/J=36$, and the comparison
  with (normalized in the vertical axis) experimental results from Ref.~\onlinecite{stoferle_tonks_optical}.}\label{fig:spectralWeightAndShape1D}
 \end{center}
\end{figure}

\begin{figure}[h]
 \begin{center}
 \includegraphics[width=\figwidth]{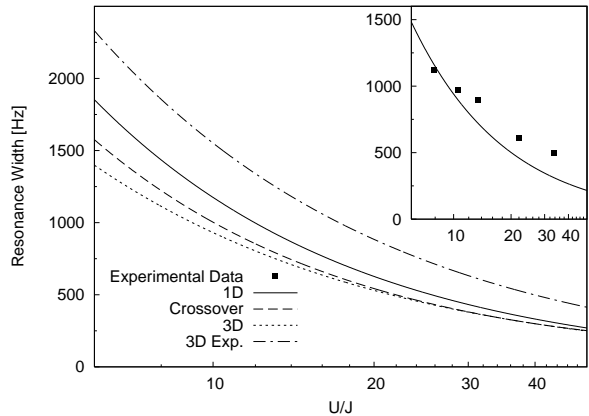}
 \caption{Width at the \emph{base} of the first resonance peak in the MI phase. In the 1D, 1D-3D crossover and
  3D cases we take $V_\perp=30E_R$, $V_\perp=20E_R$ and $V_\perp=V_{x0}$ respectively, and $n_0=1$. The 3D Exp.
  case is the same than 3D but with $n_0=2$, which is closer to the experimental values. The inset
compares the predicted \emph{half} widths with the experimental
data from Ref.~\onlinecite{stoferle_tonks_optical} for 1D.
}\label{fig:width}
 \end{center}
\end{figure}

The EAR can thus be computed by studying the linear response to
the last term of (\ref{perturbation}), via a strong coupling
expansion in $J_\alpha/U$. We sketch here the derivation in the 1D
case, full details and the more involved higher dimensional case
can be found in~\cite{iucci_absorption_long}. In the MI phase with
$n_0$ bosons per site ($n_0$ is an integer), the ground state of
$H_U$ is $|\Phi_0\rangle = | n_1 = n_0, \ldots, n_M = n_0
\rangle$, while the first excited state $|\Phi(R,r)\rangle$ with
the \emph{same} number of bosons   has an extra  boson
(``particle'') at site $R = 1,\ldots, M$ and one fewer boson
(``hole'') at  $R+r$ ($r = 1, \ldots, M-1$) and costs an energy
$U$. To obtain the linear response for $\hbar \omega \approx U$,
we only need the ground state and the states with one particle and
one hole. To take into account $H_J$, one needs to diagonalize the
kinetic energy in a subspace where the particle and the hole hop,
but no additional pair is created or destroyed.
%, and thus the
%particle and the hole cannot hop on the \emph{same site}.
Thus, at \emph{arbitrary} filling $n_0$, we employ the ansatz  $|
\Phi(Q, q) \rangle = \frac{\sqrt{2}}{M} \sum_{R,r} e^{i Q R } e^{i
\theta(Q) r} \sin \left(\frac{ q r }{2} \right) | \Phi(R, r)
\rangle$ ($ Q = 2\pi j/M$ and $q = 2\pi l/M$ for  $M$ lattice
sites with periodic boundary conditions). The eigenvalues of the
kinetic energy are $\epsilon(Q,q) = U - J\rho(Q)\cos(q/2)$. Here
$\rho(Q)$ and $\theta(Q)$ are the modulus and the argument of $n_0
+e^{i Q}(1+n_0)$. Using the spectral decomposition of the
correlation function $\chi_{\tilde {\cal O}}(\omega)$ (to leading
order in perturbation theory in $\delta J_1$),
\begin{equation}
 \dot{\varepsilon}(\omega) \approx \frac{2\pi\omega}{N}\sum_{Q,q}
 \left\vert\left\langle\Phi(Q,q)\right\vert \tilde{{\cal O}}
\left\vert\Phi_0\right\rangle\right\vert^2
 \delta\left[\hbar\omega - \epsilon(Q,q)\right],
\end{equation}
we get the EAR  in 1D to be:
\begin{equation}\label{eq:peakShape}
 \dot{\varepsilon}(\omega)=\frac{\delta J_1^2}{2J_1}
 \left(\frac{n_0+1}{2n_0+1}\right)\,\omega\,
 \sqrt{1-\left[\frac{\hbar\omega-U}{(2n_0+1)J_1}\right]^2}.
\end{equation}
There is thus a resonance at $\hbar\omega \approx U$ with a width
at the \emph{base} of the peak $2W = 2(2n_0+1)J_1$. At lowest
order in $\delta J_1$ the absorption is zero for $\hbar \omega < U
- W$ or $\hbar \omega
> U + W$.  Note that the shape of the resonance peak is not
symmetric around $\hbar\omega=U$.

This is shown in the inset of
Fig.~\ref{fig:spectralWeightAndShape1D}, where the absorption
maximum is at $\hbar\omega\approx U[1+(2n_0+1)^2J_1^2/U^2]$. The
square root form in Eq. (\ref{eq:peakShape}) is a non-perturbative
result, i.e., an effective resumation of a certain class of
diagrams \cite{iucci_absorption_long}. The peak spectral weight,
$f$, can readily be computed:
\begin{equation}
 f = \frac{\pi(n_0+1)}{4\hbar^2}\,U\,(\delta J_1)^2,
\end{equation}
and is shown in Fig.~\ref{fig:spectralWeightAndShape1D}. In $d>1$
there is no  simple formula for the precise shape of the resonance
peak. However, the width of the peak still  has the same form,
provided that $J_1$ is replaced by $J$. Moreover, it does not
depend on  $\delta J_\alpha$, and therefore it is unaffected by
the a lattice modulation in one or more dimensions. In
Fig.~\ref{fig:width}, we plot $2W$ as a function of $U/J$, for the
same setup as in Ref.~\onlinecite{stoferle_tonks_optical}. The
line shape shown in the inset of Fig. 3 is in good agreement with
the experimental observations reported in
Ref.~\onlinecite{schori_absorption_optical,stoferle_tonks_optical}.

Finally we compare our results with the experimental findings. In
the strong coupling MI phase, our results are in good agreement
with the experimental observations: (i) In 1D, the line shape
shown in the inset of Fig.~\ref{fig:spectralWeightAndShape1D} is
in good agreement with the experimental findings reported in Refs.
\onlinecite{stoferle_tonks_optical} and
\onlinecite{schori_absorption_optical}. (ii) As shown in
Fig.~\ref{fig:width}, for $U/J \gtrsim 10$, the width of the peak
at $U$ decreases with $U/J$ in all dimensions, which is again in
agreement with the experimental observation. Our values for the
half width in 1D are also in good agreement with experiments, as
shown in the inset, except for a slight
 broadening that is due to the
trapping, and also to the fact that the experiments are done in a
multiple tubes setup, which changes the filling from one tube to
another. (iii) In experiments, for fixed $U/J$, the width of the
resonance becomes \emph{smaller} as one goes from the 1D to 3D
lattices. Our calculations capture this effect in the 1D and 1D-3D
crossover cases, although a larger value of the width at the base
is predicted for 3D~\footnote{We are not able to compute the half
widths in 3D, which makes comparisons with experiments difficult.
Note also the difference in  fillings in 1D and 3D: in experiments
the average boson occupancy per site is 1.2 (1D), and 2.4  (3D)
whereas we have taken $n_0 = 1$ (1D) and $2$
(2D).\label{fn:filling}}. However we would like to emphasize that
here we compute the width at the base whereas the half width is
fitted in the experiments. Clearly, more experimental data in the
MI regime are needed to test not only for the width of the
resonance peak~\footnote{In particular, comparing an estimate of
the peak width to our linear response result can give an idea of
the importance of non-linear effects.}, but also its spectral
weight. In the weak coupling regime, as noted above, our results
are also able to explain qualitatively the broad absorption
spectra measured in the experiments for 1D systems, with a
continuous absorption curve for the SF, starting from $\omega=0$.
Accurate measurements of the low-frequency part of the spectrum
should allow, in principle, to determine the parameter $K$, which
characterizes long-range correlations in the LL.

We thank M. K\"ohl for useful correspondence on experimental data
and I. Bloch, E. Demler, A. Georges, C. J. Bolech and C. Kollath
for interesting discussions. Part of this work was supported by
the Swiss National Fund under MaNEP. MAC also acknowledges support
from \emph{Gipuzkoako Foru Aldundia} (Basque Country), and AFH
from EPSRC (UK).

\end{document}